# *Technologie et pratiques bibliographiques associées à l'écriture scientifique en milieu universitaire*

**Résumé :**
Observer et comprendre les populations d'usagers de l'information scientifique et technique, c'est se préparer à leur offrir des services adaptés. Cette étude exploratoire offre des réponses sur les usages en sciences humaines et sociales, ainsi qu'en sciences dures. Nous observerons également les personnes qui assistent les enseignants chercheurs dans leur production scientifique : les personnels de services documentaires. Nous dégagerons ensuite des réflexions et préconisations pour orienter les fonctionnalités des services numériques à destination des acteurs de la recherche de littérature scientifique.

## 1. Introduction

Nous ambitionnons de modéliser un système de recherche d'informations (SRI) scientifique pour les enseignants chercheurs, les étudiants de troisième cycle et les personnels spécialisés en documentation qui les aident dans leur phase de recherche documentaire au sein des services communs de documentation (SCD[1]). La présente étude vise à réunir de l'information autour des pratiques documentaires dans l'optique d'en faire émerger un cahier des charges afin d'optimiser les fonctionnalités du SRI. En effet, placer les pratiques des usagers comme fondement de la conceptualisation d'un système d'information (SI), c'est créer un SI à « valeur ajoutée globalisée » [1]. Selon de Kaenel et Iriarte, « les dernières évolutions du web, avec l'entrée en jeu de XML, des nouveaux usages et nouveaux outils, ainsi que le déplacement du centre de gravité qui s'est fortement rapproché des utilisateurs, ouvrent de nouvelles voies et de nouveaux champs d'application pour les catalogues en ligne » [2].

Nous organiserons notre étude autour de plusieurs axes techniques et sociologiques. L'axe principal est technique, il se propose de faire émerger les protocoles de recherche des universitaires (doctorants, documentalistes et enseignants chercheurs). Ces protocoles seront établis sociologiquement en fonction du niveau d'expertise des usagers et de leurs centres d'intérêts scientifiques.

Jusqu'en 2005, nous ne pouvions que nous étonner du peu d'études françaises récentes qui s'interrogeaient sur les pratiques des usagers des systèmes numériques de recherche d'informations, notamment scientifiques. En effet, comment modéliser un système de recherche d'informations (SRI) tout en ignorant le contexte d'usage ? Placer l'humain au centre du processus de modélisation permet de proposer un outil au plus près des besoins et attentes des usagers. Depuis quelques années, l'étude des usages regagne du terrain plus particulièrement en sciences dures principalement grâce à A.

---

[1] Un Service Commun de la Documentation rassemble les bibliothèques d'un établissement de l'enseignement supérieur.

Wojciechowska en informatique, mathématique (Maths-Info) [3], J. Schöpfel et C. Boukacem-Zeghmouri en sciences, techniques et médecine (STM) [4]. Nous élargirons notre enquête aux sciences humaines et sociales (SHS).

Des études précédentes synthétisaient également la littérature sur la méthodologie de recherche d'informations sous des aspects psycho cognitifs [5][6]. Carol Kuhlthau mettait en évidence les étapes du processus de recherche d'informations tout en y associant les sentiments, pensées de l'usager. Si l'utilisateur du SRI ne trouve pas rapidement l'information qu'il cherche lors des différentes étapes du processus de recherche, il va très vite se décourager et être plongé dans un état d'insatisfaction qui le fera finalement renoncer [7].

Nous pensons que cet état de satisfaction peut être en corrélation, bien entendu avec la qualité des métadonnées recueillies, mais aussi avec la compatibilité entre le SRI et les outils que l'usager s'est choisis. Ainsi, l'usager incapable de charger de manière simple et intuitive le résultat de sa recherche – à savoir une ou plusieurs notices bibliographiques — se retrouverait dans un état d'insatisfaction qui le pousserait à changer de SRI.

Ainsi, les aspects simplement quantitatifs et qualitatifs en terme de données fournies par le SRI, s'ils sont indispensables pour créer un système de recherche d'informations, ne sont pas suffisants pour le rendre acceptable par les usagers.

Bermès et Martin déclaraient qu'une « collection numérique ne peut pas être appréhendée directement. Elle requiert une médiation technique (…) entre l'usager et la collection » [8]. L'objet de la présente étude est précisément de déterminer des profils d'usagers de l'information scientifique et technique (IST) au travers de l'observation de leurs choix techniques. Ainsi, munis de précieuses informations relatives aux usages, nous tenterons de synthétiser les besoins des utilisateurs en terme de « médiation technique ». Ces données permettront ultérieurement de créer un outil répondant aux besoins en terme d'usage technique par une urbanisation du SI *had hoc* [9]. Nous entendons par ce terme d'urbanisation les fonctionnalités communicationnelles entre « briques logicielles » d'un système d'informations. Cela se concrétise par des fonctions *Extract Transform Load* capables de fournir le flux informationnel dans le format attendu [10] par un processus automatisé. Il n'est en effet pas possible de modéliser un SRI et plus généralement un SI[2] sans se référer directement aux technologies de l'information [12].

Pour les doctorants et les enseignants chercheurs principalement, nous avons tenu à distinguer principalement les sciences dites « dures », des sciences humaines et sociales. Cette distinction est importante en terme d'usages, car les outils et protocoles de production scientifique diffèrent. En effet, en sciences dures, il est courant que la demande des revues et conférence impose un protocole d'écriture précis. Les documents doivent ainsi parfois être rédigés en utilisant le format LaTeX, ce qui induit l'usage du format bibliographique BibTeX. Nous allons tenter de spécifier les pratiques autour de ces contraintes par la présente enquête. En sciences humaines et sociales, les usages diffèrent, nous allons en définir la mesure.

---

[2] Définition de Système d'information « Processus qui collectent des données structurées conformément aux besoins (…), qui stockent, traitent et distribuent l'information » [11]

Les différences ne s'arrêtent pas à la production normée de documents, les usages en terme d'outils varient également, notamment en matière de recherche d'informations et de stockage de notices bibliographiques.

## 2. Résultats attendus

En théorie, cette étude devrait permettre de spécifier des profils donnant lieu à des préconisations d'usage, utiles surtout aux novices.

D'autres études précédentes donnaient une idée des logiciels et formats les plus usités pour la gestion bibliographique. Une étude suisse de 2009 présente les logiciels proposés par les SCD des universités suisses [13]. Masur concluait que le public universitaire connaissait encore très mal en 2009 les LGRB[3], que les outils libres peinaient à se faire une place au sein de l'offre. Masur terminait sur la prédiction que les LGRB prendront de l'importance dans un proche avenir. Nous nous baserons principalement, pour formuler nos questions en ce domaine, sur l'étude de Carole Zweifel qui présente 7 des principaux logiciels de gestion bibliographique [14]. Nous avons sélectionné parmi ces logiciels ceux dits autonomes (qui ne nécessitent pas l'installation d'un serveur hypertexte et d'une base de données en pré requis). Nous avons rajouté RefBase qui est parfois installé par les services informatiques universitaires comme outil de partage bibliographique pour une équipe ou un laboratoire. Nous avons complété notre panel avec les deux outils propriétaires parmi les plus connus : EndNote et RefWorks. Nous offrons donc au sondé de choisir le ou les outils de gestion bibliographique qu'il utilise le plus parmi 6 des plus connus. Il a également la possibilité de proposer un outil alternatif s'il le désire.

Il en va de même pour les sources d'informations scientifiques et techniques dans le milieu de la recherche. Nous allons réactualiser et mettre en contexte ces données.

Mais également et surtout, nous obtiendrons des informations précieuses sur les caractéristiques techniques de compatibilité que doit proposer un système de recherche d'informations en adéquation avec ses usagers. De plus, cette étude pourra être utilisée pour mesurer l'adéquation technique des SRI en information scientifique et technique avec les outils communément usités par les communautés de chercheurs.

Dans un premier temps, nous allons proposer les éléments de notre questionnaire et les objectifs induits par leur recoupement.

### Méthodologie d'enquête

Pour encourager les utilisateurs à répondre au questionnaire, nous avons volontairement simplifié le questionnaire en réduisant la typologie des questions. Ainsi, le temps des sondés est épargné au maximum. Nous ne ferons pas intervenir ici de questions sur l'âge ou le sexe, ce qui nous semble sans objet, voire déplacé.

### A) Typologie des questions
- Questions fermées : 1 choix parmi 3 ou 4.
- Questions semi-fermées : 1 ou plusieurs choix parmi plusieurs.
- Questions semi-ouvertes : 1 ou plusieurs choix parmi plusieurs ou la possibilité de fournir sa propre réponse.

---

[3] Acronyme de Logiciels de Gestion de Références Bibliographiques, imputé à Zweifel

Nous avons établi le plan de sondage de la manière suivante :

1. Questionnaire filtrant pour établir des profils scientifiques avec des questions fermées (statut, expérience, science étudiée).
2. Questionnaire fermé sur le contexte technique d'écriture scientifique (système d'exploitation et logiciels connus et utilisés).
3. Questionnaire plus spécifique sur les pratiques bibliographiques avec des questions fermées et semi-ouvertes.
4. Questionnaire fermé et semi-ouvert sur l'impact des aspects de logiciels libres et de gratuité sur le choix de l'utilisateur.
5. Questionnaire des priorités dans le choix d'un logiciel bibliographique.
6. Question semi-ouverte sur le choix des sources d'informations scientifiques.

**b) Population cible et Panel**

La question de la représentativité d'un panel de sondés est cruciale pour la crédibilité d'une enquête.

À partir d'une vingtaine d'observations (par échantillon de la population observée), des tendances émergent. Cependant, plus le public est hétérogène, plus il faut élargir le panel [15]. Notre étude, extrêmement ciblée, est principalement quantitative. Dans notre cadre, contrairement aux sondages d'opinion, il n'est pas nécessaire d'interroger beaucoup d'individus de chaque catégorie. Pour chaque population étudiée, 20 à 25 personnes suffisent pour obtenir de bons résultats [16]. Quand à la marge d'erreur, pour une population entre 100 et 200 personnes, dans le cadre de ce type de sondage, elle se situe entre 3.1 et 4.4 points si l'on se réfère au tableau d'intervalle de confiance [17]. Cette marge d'erreur semble acceptable.

Cette enquête a été menée sur une période de 15 jours en décembre 2011. Ce formulaire proposé sur la base du bénévolat et de l'anonymat proposait aux usagers de recevoir un compte-rendu du rapport après publication des résultats, s'ils laissaient leur courriel. Notre enquête a été proposée en ligne sur Google Formulaires et soumise par listes de diffusion :

- à la liste de diffusion de l'association EGC[4] ;
- à l'école doctorale Cognition Langage Interaction de l'Université Paris 8 ;
- aux laboratoires d'informatiques LIAFA et LIASD respectivement rattachés aux Universités Paris 7 et de Paris 8.
- à de grands établissements comme Sciences Po et à l'ensemble des doctorants de l'école doctorale Abbé Grégoire du CNAM ;
- aux conservateurs et bibliothécaires des universités parisiennes ;
- à la liste de diffusion de l'ADBS.

Nous évaluons le nombre de personnes contactées à 9000 personnes entre les grosses listes de diffusion (ADBS : 7400 et EGC : 1000) et les petites listes (600). Le panel de personnes ayant répondu est constitué de 195 participants soit 2,17 %, dont la distribution est la suivante :

- 54 enseignants chercheurs dont 16 en sciences dures, 36 en sciences humaines et sociales et 2 dans d'autres domaines.

---

[4] EGC - Association Extraction et Gestion des Connaissances: http://www.egc.asso.fr/

- 98 doctorants dont 11 en sciences dures et 79 en sciences humaines et sociales, 8 dans d'autres domaines.
- 28 personnels de SCD
- 4 post doctorants
- 11 personnels hors de ce classement comme des ingénieurs de recherche.

|  | Doctorant début de thèse | Doctorant fin de thèse |
|---|---|---|
| Sciences humaines et sociales | 34 | 45 |
| Sciences dures | 5 | 6 |
| Autres | 5 | 3 |

Tableau 1. Répartition des doctorants par expérience et domaine de recherche

Nous n'étudierons pas en détail le statut de post doctorant, car le segment interrogé ne présente pas un corps suffisamment élevé pour être représentatif. Nous avons segmenté le statut de doctorant, pour spécifier l'expérience, en début (44 personnes) et fin de thèse (54 personnes). Pour spécifier les profils utilisateurs, nous avons encore dégagé 4 sous-ensembles principaux en fonction de l'expérience et du domaine de recherche :

Le tableau 1 montre qu'il sera pertinent de faire une distinction sur l'évolution des usages technologiques liés aux pratiques bibliographiques pour les doctorants entre le début et la fin de la thèse pour les sciences humaines et sociales.

Passons à l'examen des résultats de l'enquête.

## Les résultats
Le dépouillement des résultats en matière de données a donné les résultats suivants :

### a) Contexte technologique
Les sondés ont présenté leur contexte d'usage de l'outil numérique au travers de quelques questions. Ces questions portent sur l'environnement technologique entourant leur production scientifique. À la question du système d'exploitation favori, les chiffres de la population sont :

| Système d'exploitation | % de la population totale | SHS % du segment | Sc. Dures % du segment | Autres et sans objet % du segment |
|---|---|---|---|---|
| Linux | 15 % | 7,8 % | 36,21 % | 0 % |
| Windows | 63 % | 70,43 % | 39,65 % | 83,36 % |
| Mac OS | 22 % | 21,74 % | 24,14 % | 13,64 % |
| Autre | 0 % | 0 % | 0 % | 0 % |

Tableau 2. Contexte technologique des usagers, selon la répartition scientifique.

De manière générale, sans plus spécifier la population, il apparaît que de manière générale, un peu plus de 20 % des sondés utilisent des ordinateurs de marque Apple, que ce soit en sciences humaines et sociales ou en sciences dures. Les systèmes d'exploitation de Microsoft sont utilisés par plus des deux tiers des usagers de SHS. Le pourcentage d'usage de Mac 0S est le même en SHS que pour l'ensemble de la population de notre enquête, à savoir environ 22 %. Avec 8 %, l'usage de systèmes Linux reste marginal en SHS.

En sciences dures, la répartition est plus équilibrée entre Linux (36 %) et Windows (40 %). Mac 0S est une fois de plus utilisé par environ un quart du segment. Il est à noter que le système Mac OS X est basé sur un ancêtre commun à Linux et peut être également être utilisé comme tel [18].

Comme les chiffres montrent que seules les sciences dures ont un réel impact sur le choix d'un système d'exploitation, nous allons affiner notre analyse par type d'usagers (Doctorants, enseignants chercheurs et documentalistes).

| Systèmes utilisés en Sc. dures | Doctorants en début de thèse | Doctorants en fin de thèse | Post doctorants | Enseignants chercheurs | Documentalistes |
|---|---|---|---|---|---|
| Linux | 40 % | 66 % | 50 % | 36,21 % | 0 % |
| Windows | 60 % | 0 % | 0 % | 39,65 % | 100 % |
| Mac OS | 0 % | 33 % | 50 % | 24,14 % | 0 % |

Tableau 3. Systèmes d'exploitation utilisés par profil utilisateurs en sciences dures.

Les résultats du tableau 3 montrent qu'en sciences dures, les chercheurs débutants utilisent Windows puis glissent vers l'utilisation de Linux ou Mac (qui est basé sur Linux) en fin de thèse. Les enseignants-chercheurs retourneront à 40 % sur système Microsoft, 36 % resteront sous Linux et 14 % sous Mac.

En début de thèse, le jeune chercheur de sciences dures utilise Windows (60 %) et Linux (40 %). Il apparaît très nettement que les postes de travail sont équipés avec un système Linux ou compatible (Mac) dès la fin de la thèse et pour les populations de chercheurs.

Les documentalistes et bibliothécaires de sciences dures semblent utiliser exclusivement des systèmes Microsoft.

En SHS, la proportion d'usages diffère notablement. L'utilisation de Linux est anecdotique chez les doctorants et marginale chez les Enseignants chercheurs. Ce qui est surprenant, c'est que plus de 15 % des documentalistes spécialisés en SHS utilisent Linux. De plus, plus des trois quarts de ce segment de la population utilisent Mac OS. Ce système d'exploitation semble être particulièrement apprécié par les jeunes docteurs et les enseignants chercheurs. Windows a plus les faveurs des doctorants et d'une bonne partie des enseignants chercheurs.

| Systèmes utilisés en SHS | Doctorants en début de thèse | Doctorants en fin de thèse | Post doctorants | Enseignants chercheurs | Documentalistes |
|---|---|---|---|---|---|
| Linux | 3 % | 9 % | 0 % | 12,50 % | 15,4 % |
| Windows | 85 % | 71 % | 50 % | 43,75 % | 7,7 % |
| Mac OS | 12 % | 20 % | 50 % | 43,75 % | 76,9 % |

Tableau 4. Systèmes d'exploitation utilisés par profil utilisateurs en sciences humaines et sociales.

### b) Usages d'outils de productions écrites en sciences

Examinons quels sont les outils utilisés pour l'écriture de littérature scientifique par notre population.

Les données brutes d'utilisations d'un éditeur graphique (Word ou OpenOffice…) ou d'un compilateur de texte[5] classent l'utilisation de la manière suivante :

1. Éditeur graphique : 149
2. Compilateur de texte : 44
3. Autre : 2

Analysons par répartition ces données.

| Outil de production de documents | Doctorants début de thèse | Doctorants en fin de thèse | Post doctorants | Enseignants chercheurs | Personnels de SCD | Autres Ingénieurs |
|---|---|---|---|---|---|---|
| TeX ou LaTeX | 10 % | 18,5 % | 50 % | 46,3 % | 3,6 % | 18,2 % |
| Éditeur graphique | 90 % | 81,5 % | 50 % | 53,7 % | 92,8 % | 71,7 % |
| Autre | 0 % | 0 % | 0 % | 0 % | 3,6 % | 9,1 % |

Tableau 5. Outil de production de documents par profil d'utilisateurs.

De manière générale, les chiffres montrent que, dans le cadre scientifique, l'usage des éditeurs graphiques comme Word et OpenOffice est largement privilégié. Cependant pour le segment des enseignants chercheurs, l'écart entre le pourcentage d'usagers de compilateurs de texte et de traitement de texte est très réduit. Dans les autres groupes, l'usage des compilateurs de texte est quasi nul, sauf pour les doctorants en fin de thèse et les ingénieurs (18 % dans les deux cas).

En sciences humaines, l'usage du compilateur évolue avec l'expérience de presque nul, 3 % en début de thèse à 12,5 %, un usage très modéré pour les enseignants chercheurs.

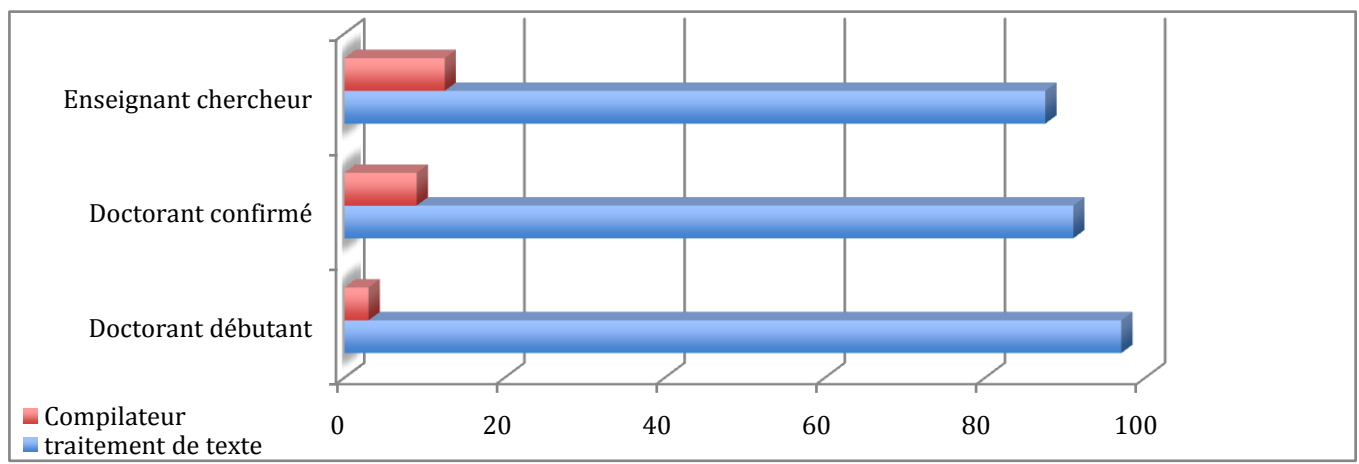

Figure 1. Méthode d'écriture en SHS selon l'expérience

L'usage d'un compilateur est beaucoup moins anecdotique en sciences dures, notamment pour les doctorants en fin de thèse. Cette partie de la population rédige son mémoire de thèse en sciences dures, avec des équations et une bibliographie complexes. Entre ces considérations techniques et des pratiques sociologiques universitaires très orientées vers LaTeX en sciences dures, l'usage d'un compilateur de texte est presque une thèse de doctorat.

---

[5] Un compilateur de texte est un logiciel qui génère un document formaté portable (PDF) à partir d'un langage de mise en page de texte comme TeX ou LaTeX.

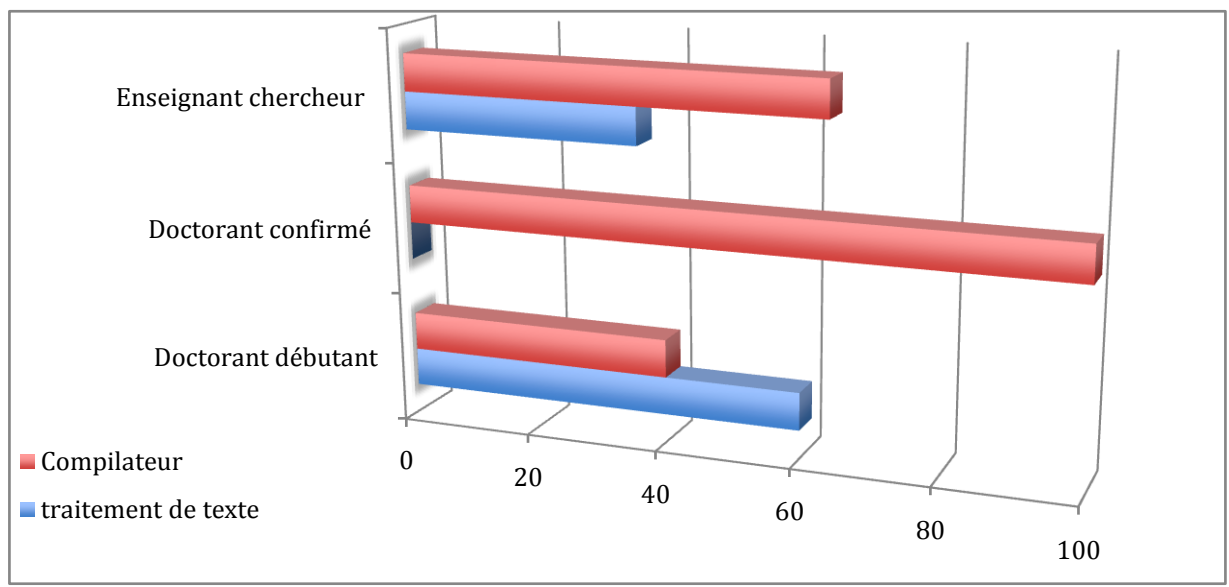

**Figure 2. Méthode d'écriture en sciences dures**

Chez les documentalistes et bibliothécaires, il est plutôt courant d'utiliser un traitement de texte (97 %). Les 3 % restants sont les compilateurs et autres.

Si nous croisons les informations relatives à l'usage d'un compilateur de texte avec celles concernant le système d'exploitation utilisé, nous avons la surprise de constater que son usage n'est pas lié à un système d'exploitation en particulier :

Les enseignants chercheurs (tous issus de sciences dures) usants de LaTeX ou TeX utilisent les 3 systèmes d'exploitation avec une majorité de Linux.

|         | Doctorants début de thèse | Doctorants en fin de thèse | Post doctorants | Enseignants chercheurs | Personnels de SCD | Autres Ingénieurs |
|---------|---------------------------|----------------------------|-----------------|------------------------|-------------------|-------------------|
| Windows | 0 %                       | 10 %                       | 0 %             | 20 %                   | 0 %               | 100 %             |
| Mac OS  | 50 %                      | 30 %                       | 50 %            | 32 %                   | 0 %               | 0 %               |
| Linux   | 50 %                      | 60 %                       | 50 %            | 48 %                   | 100 %             | 0 %               |

**Tableau 6. OS en fonction de l'usage d'un compilateur de texte (LaTeX, TeX).**

Le tableau 6 illustre le fait que les usagers de LaTeX utilisent principalement un système compatible Unix, à savoir Linux ou Mac 0S.

Cette population édite les fichiers LaTeX de la manière suivante :

- Éditeur à compilation intégrée (TexWorks, Emacs…) : 72 %
- Éditeur basique (vi, ed, Notepad…) : 15 %
- autre : 22 %

Les propositions pour les utilisateurs de TeX/LaTeX, ayant coché la valeur « autre » ont été : Kyle[6], 4 personnes ont proposé LyX[7] dont 3 en sciences humaines, TeXnicCenter[8], usbTex[9] et Eclipse avec le plug-in TeXlipse.

**c) Les usages et formats bibliographiques**

Pour mieux cerner les usages scientifiques en matière de bibliographie, nous avons demandé à la population de sondés s'ils intégraient leurs citations et bibliographies manuellement où s'ils la généraient grâce au compilateur de leur outil de traitement de texte.

Il semble que de manière globale, l'introduction des références et bibliographies se fasse encore beaucoup manuellement dans le texte (37 %). Pour ceux qui utilisent un fichier séparé, ils sont 19 % à l'enrichir manuellement et 18 % à utiliser un logiciel bibliographique. En tenant compte des 25 % de personnes qui utilisent l'outil interne à leur traitement de texte, 63 % des sondés en milieu universitaire utilisent une fonction automatisée de gestion de bibliographie. Ce chiffre mérite d'approfondir les pratiques liées à ces outils pour mieux définir les services que doivent rendre les SRI, pour être en adéquation avec les besoins des usagers.

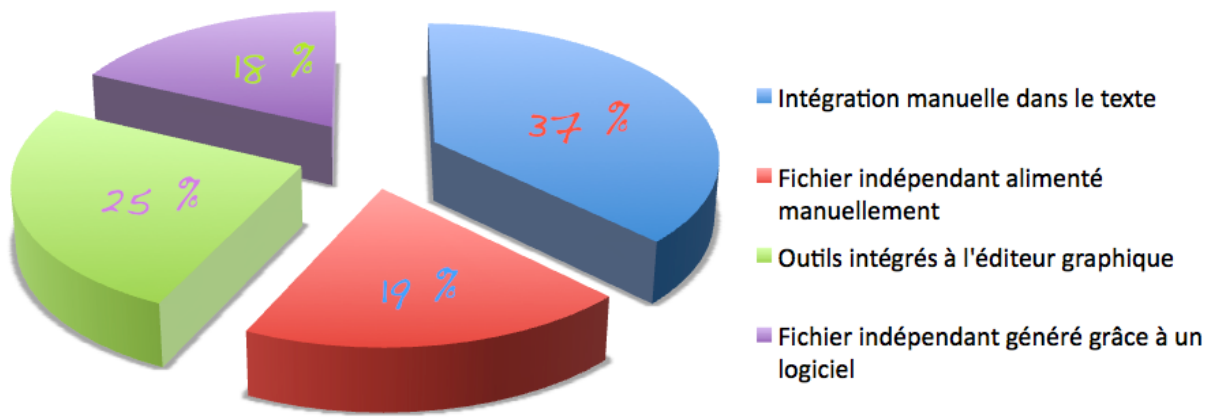

Figure 3. Méthode d'intégration bibliographique dans un document pour l'ensemble de la population.

Si l'on répartit ces résultats par sciences, 20 % des sondés en sciences dures intègrent manuellement leurs références dans le texte pour 47 % en sciences humaines. La population la plus éloignée de cet usage est celle des employés de SCD puisque seuls 17,86 % de ceux interrogés intègrent leurs références manuellement dans le texte.

Nous avons questionné les usagers sur les formats de fichiers bibliographiques qu'ils connaissent et ceux qu'ils utilisent. Nous voyons, grâce à la figure 2 que 39 % des sondés ne connaissent aucun des formats bibliographiques présentés. Il s'agit principalement des chercheurs et personnels en sciences humaines et sociales et autres. En sciences dures, seules 3 personnes (soit 5 %) ne connaissent aucun format de fichiers bibliographiques. Il s'agit d'usagers de Windows avec des éditeurs graphiques, comme Word ou OpenOffice. Il est à noter que seuls 41,7 % des sondés issus des SHS connaissent au moins un format de fichier bibliographique parmi ceux que nous avons proposés.

---

[6] Kile est un éditeur graphiqe de TeX / LaTeX disponible sur Mac OS, Linux et Microsoft Windows : http://kile.sourceforge.net/ et http://www.framasoft.net/article2827.html
[7] Environnement TeX avec une interface entièrement graphique utilisable à la souris http://www.framasoft.net/article1001.html
[8] Environnement intégré de développement TeX en C++ sous Windows http://www.framasoft.net/article1429.html
[9] Environnement TeX complet transportable sur clé USB : http://www.framasoft.net/article4641.html

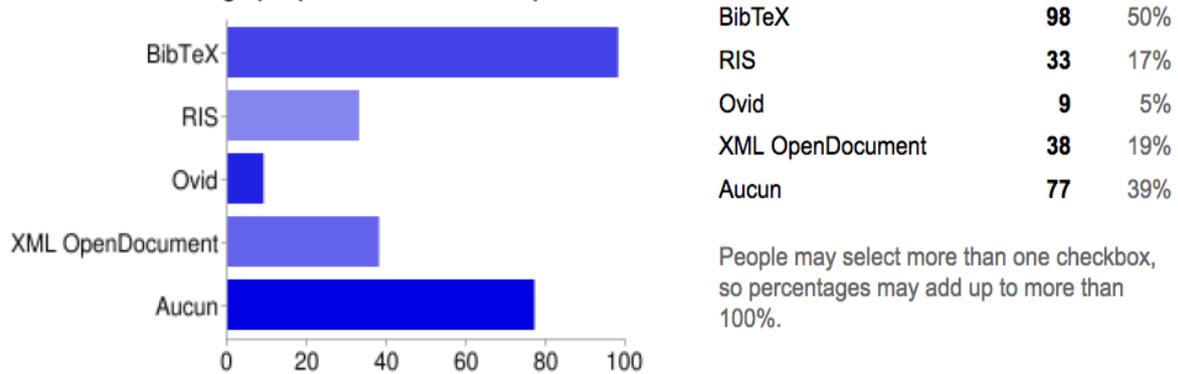

Figure 4. Formats de fichiers bibliographiques connus par l'ensemble de la population.

En sciences humaines toujours, nous allons examiner la connaissance de ces formats de fichiers bibliographique par les sondés de notre échantillon. Le tableau 7 se lit de la manière suivante :

- Le terme « *connaissance unique* » signifie que le sondé connaît ce format bibliographique et seulement celui-là.
- Le terme « *connaissance multiple* » signifie que le sondé connaît ce format parmi d'autres.
- Les valeurs en rouge indiquent le pourcentage de la population en SHS qui connaît le format bibliographique.
- Les valeurs en vert indiquent le pourcentage de personnes connaissant ce format bibliographique au sein du sous-ensemble composé des membres de la population de SHS qui connaissent au moins un format bibliographique.

| Formats connus par la population SHS | BibTeX | RIS | Ovid | XML OpenDocument | Aucun |
|---|---|---|---|---|---|
| Connaissance unique | 11,3 % | 2,6 % | 1,7 % | 8,7 % | 58,26 % |
| Connaissance multiple | 16,5 % | 5,2 % | 2,6 % | 8,7 % | - |
| % de la population SHS | 27,83 % | 7,83 % | 4,35 % | 17,39 % | 58,26 % |
| % parmi la population SHS connaissant au moins un format bibliographique | 66,67 % | 18,75 % | 10,42 % | 41,67 % | - |

Tableau 7. Formats bibliographiques connus en SHS

Le format le plus connu en SHS est le BibTeX avec presque 28 % des sondés. Le format bibliographique OpenDocument compatible avec Word et OpenOffice n'est connu que par moins de 20 % du segment. Le RIS et l'Ovid sont connus très à la marge, presque de manière anecdotique. Cela s'explique bien pour l'Ovid qui est un format dédié aux sciences médicales. Voyons si ces résultats sont comparables en sciences dures.

| Formats connus par la population de sciences dures | BibTeX | RIS | Ovid | XML OpenDocument | Aucun |
|---|---|---|---|---|---|
| Connaissance unique | 60,3 % | 0 % | 1,7 % | 1,7 % | 5,2 % |
| Connaissance multiple | 31 % | 15,5 % | 3,4 % | 20,7 % | - |
| % de la population en sc. dures | 91,38 % | 15,52 % | 5,17 % | 22,41 % | 5,17 % |
| % de la population en sc. dures connaissant au moins un format. | 96,36 % | 16,36 % | 5,45 % | 23,64 % | - |

Tableau 8. Formats bibliographiques connus en Sciences dures

En sciences dures, l'écrasante majorité de l'échantillon connaît le format BibTeX et une proportion appréciable (22,41 %) connaît le XML OpenDocument. Notons que la proportion de la population connaissant au moins un format tend à rejoindre le chiffre de la population, puisque seuls 5,17 % des sondés de sciences dures ne connaissent aucun format. Aucun doctorant ou post doctorant ne se trouvait dans cette situation et seuls 5,5 % des enseignants chercheurs de sciences dures ne connaissant aucun format.

Pour ce qui est des bibliothécaires et documentalistes sondés, nous les avons placés à part. Leur statut leur permettant de se positionner au plus près des documents scientifiques, ils sont généralement au fait des normes et pratiques bibliographiques universitaires. Nous avons même constaté que certains connaissaient tous les formats présentés.

Comme prévu, les documentalistes et bibliothécaires sont particulièrement sensibles aux formats bibliographiques. Une fois de plus, le format BibTeX est le plus connu avec plus des deux tiers des sondés qui le connaissent. Le format RIS est également connu par plus de la moitié des personnels de SCD interrogés. Le format de bibliographie d'OpenDocument est connu de presque la moitié de ce panel. Dans l'ensemble, les personnels de bibliothèques ont la connaissance la plus large (pas uniquement le BibTeX) de l'ensemble de la population.

| Formats connus par les bibliothécaires et documentalistes | BibTeX | RIS | Ovid | XML OpenDocument | Aucun |
|---|---|---|---|---|---|
| Connaissance unique | 3,6 % | 6 % | 6 % | 10,7 % | 14,3 % |
| Connaissance multiple | 64,3 % | 53 % | 10,7 % | 35,7 % | - |
| % de la population en SCD | 67,86 % | 53,57 % | 14,29 % | 46,43 % | 14,29 % |
| % de ce segment connaissant au moins un format | 79,17 % | 62,50 % | 17,67 % | 54,17 % | - |

Tableau 9. Formats bibliographiques connus par les bibliothécaires et documentalistes.

Notons que toutes disciplines et tous corps confondus, la moitié des sondés connaissaient le format BibTeX. Nous avons ensuite interrogé la population sur les formats de bibliographie utilisés dans le cadre de la rédaction de documents scientifiques, thèse ou de rapports techniques. Plusieurs réponses à la question étaient admises, de plus, les sondés pouvaient proposer une autre réponse s'ils le souhaitaient.

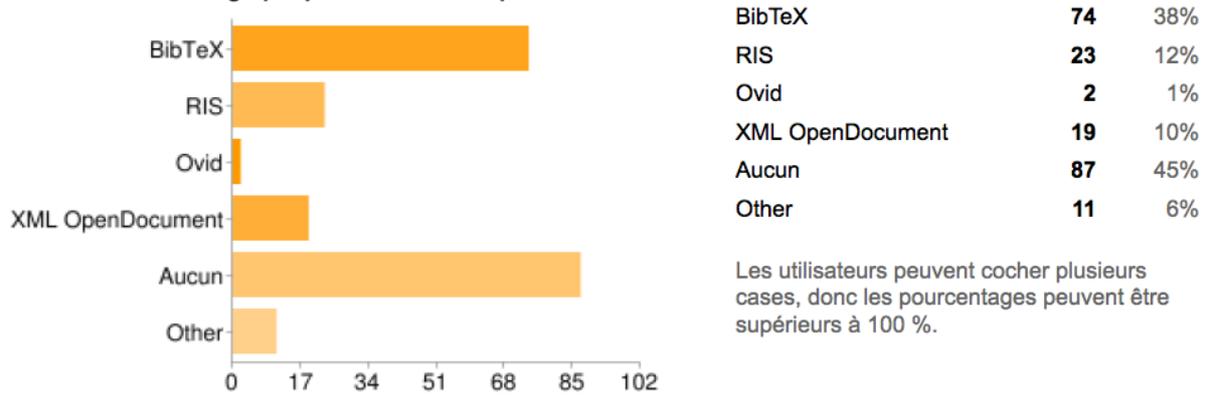

Figure 5. Formats de fichiers bibliographiques utilisés

L'analyse des chiffres avant répartition indique que le format bibliographique le plus utilisé est le BibTeX. Cependant, ce que nous apprend la figure 3, c'est principalement que 45 % de l'échantillon n'utilise aucun fichier pour gérer les bibliographies. Nous avons également eu des réponses alternatives très pointues comme : Zotero RDF[10], APA[11] (2 chercheurs), MARC[12] (une documentaliste).

En SHS, presque les deux tiers des sondés (61,74 %) n'utilisent pas de fichiers formatés pour la gestion de leur bibliographie. Parmi ceux qui n'utilisent pas de fichiers bibliographiques, 29,58 % se servent néanmoins des fonctions bibliographiques intégrées à leur traitement de texte. Pour ceux qui utilisent un fichier, c'est principalement BibTeX qui a leur faveur à presque 50 %. Pour le reste, 23,91 % des utilisateurs de fichiers bibliographiques formatés en SHS utilisent le format XML OpenDocument et presque 20 % le RIS. À la marge, 9 % des personnes interrogées, dans ce segment, utilisent d'autres formats.

| Formats utilisés par la population SHS | BibTeX | RIS | Ovid | XML OpenDocument | Autre | NSP | Aucun |
|---|---|---|---|---|---|---|---|
| Utilisation unique | 13 % | 5,2 % | 0 % | 6,9 % | 2,6 % | - | - |
| Utilisation multiple | 5,2 % | 3,5 % | 0 % | 2,6 % | 0,9 % | - | - |
| % de la population SHS | 18,26 % | 8,69 % | 0 % | 9,56 % | 3,48 % | 1,5 % | 61,74 % |
| % de la population SHS utilisant au moins un format | 47,73 % | 18,75 % | 0 % | 23,91 % | 9,09 % | - | - |

Tableau 10. Formats bibliographiques utilisés en SHS

Parmi les doctorants en début de thèse de SHS, sur une population de 34 personnes, 14,7 % utilisent le format BibTeX, ce qui est étonnamment élevé, si l'on prend en compte que beaucoup n'ont pas encore commencé la rédaction de leur manuscrit et que ce format est plus facilement utilisé en sciences dures. 3,22 % de cette même population utilise le format RIS et seulement 5,88 % utilisent le format OpenDocument, plus adapté aux éditeurs graphiques, notamment Word depuis sa version 2007.

---

[10] Format XML de description de notices bibliographiques généré pas Zotero.
[11] L'APA est un format de présentation pour les références bibliographiques, notes de bas de page, citations dans les publications. Ce format est défini par *l'American Psychological Association* [19].
[12] Format de données bibliographiques de la Bibliothèque du congrès.
http://www.loc.gov/marc/bibliographic/ (accédé le 3 janvier 2012)

Pour les doctorants de SHS en fin de thèse, sur une population de 45 personnes, 30 n'utilisent pas de format bibliographique, soit les deux tiers. Cela peut s'expliquer en regardant le tableau 5. Plus de 80 % des doctorants rédigent sur éditeurs graphiques. Avec Word par exemple, il est possible d'introduire manuellement les centaines de références bibliographiques d'une thèse et d'y faire référence.

Pour les enseignants chercheurs et post doctorants en SHS, l'échantillon s'élève à 18 individus dont 10 (soit 55,56 %) n'utilisent aucun fichier formaté de bibliographie. Pour ceux qui en font usage, 22,22 % choisissent le BibTeX, 11,11 % le RIS et 11,11 % l'XML OpenDocument.

| Formats utilisés par la population Sciences dures | BibTeX | RIS | Ovid | XML OpenDocument | Aucun |
|---|---|---|---|---|---|
| Utilisation unique | 72,4 % | 3,4 % | 1,7 % | 0 % | - |
| Utilisation multiple | 6,9 % | 5,2 % | 1,7 % | 6,9 % | - |
| % de la population Sc. dures | 79,31 % | 8,62 % | 3,45 % | 6,90 % | 13,79 % |
| % de la population Sc. dures utilisant au moins un format | 92 % | 10 % | 4 % | 8 % | - |

Tableau 11. Formats bibliographiques utilisés en Sciences dures

En sciences dures, l'usage d'un fichier séparé pour la bibliographie est de manière générale mieux ancré dans la culture scientifique. Seuls 13,8 % des sondés de ce groupe n'utilisent aucun fichier bibliographique, cependant dans ce sous-groupe seul un utilisateur sur 4 intègre ses citations manuellement, les autres utilisent les fonctionnalités prévues à cet effet par leur traitement de texte. Cependant, en sciences dures le format BibTeX est quasi exclusivement utilisé. 72, 41 % de la totalité du groupe l'utilisent exclusivement et 92 % de ceux qui utilisent au moins un format de fichier en font usage. Seul le format RIS est également utilisé comme format unique par 5 % de la population. Nous pouvons remarquer un usage marginal du XML OpenDocument (4 % des sondés en sciences dures) comme format secondaire.

| Formats utilisés par la population de personnels de SCD | BibTeX | RIS | Ovid | XML Open Document | Autre | Aucun |
|---|---|---|---|---|---|---|
| Utilisation unique | 7,1 % | 25 % | 0 % | 7,1 % | 3,6 % | - |
| Utilisation multiple | 21,4 % | 21,4 % | 7,1 % | 17,9 % | 0 % | - |
| % de la population de SCD | 28,57 % | 46,43 % | 7,14 % | 25 % | 3,57 % | 28,57 % |
| % de la population en SCD utilisant au moins un format | 40 % | 65 % | 10 % | 35 % | 5 % | - |

Tableau 12. Formats bibliographiques utilisés par les documentalistes et bibliothécaires en SCD.

Les documentalistes et bibliothécaires utilisent un large panel de format de fichiers bibliographiques. Parmi les personnes qui n'utilisent pas de fichier, 65 % utilisent le RIS et 40 % LaTeX. Le format XML est utilisé également par 35 % des utilisateurs de fichiers formatés de bibliographie. Ceux qui n'en utilisent pas se servent des outils bibliographiques intégrés à leur traitement de texte dans 75 % des cas. Sur l'ensemble de la population de SCD de notre panel, seulement 7,14 % des sondés intègrent leurs références bibliographiques manuellement dans les écrits.

### d) Les usages de logiciels de gestion bibliographique (LGRB)

Nous allons nous intéresser à l'usage qui est fait des outils qui permettent de générer et de gérer des bibliographies. Dans notre panel, 56 % en utilisent un, 38 % n'en utilisent pas et pour 6 %, cette question est sans objet.

En SHS, la proportion est la suivante :

- utilise au moins un logiciel de gestion bibliographique : 53,91 %
- n'utilise aucun logiciel de gestion bibliographique : 38,26 %
- sans objet : 7,83 %

En sciences humaines et sociales, les résultats sont presque identiques à ceux des sciences dures. Seules les pratiques liées à la non-utilisation sont distinctes. En SHS, ceux qui n'utilisent pas de logiciel dédié pour la gestion de leur bibliographie utilisent plus volontiers les fonctionnalités de leur traitement de texte (20,45 %), mais plus souvent, ils intègrent les éléments bibliographiques sans assistance logicielle (75 %) dans les documents. Les autres entretiennent un fichier bibliographique manuellement.

En Sciences dures, les chiffres sont les suivants :

- utilise au moins un logiciel de gestion bibliographique : 53,45 %
- n'utilise aucun logiciel de gestion bibliographique : 41,38 %
- sans objet : 5,17 %

De manière générale, en sciences dures, ceux qui n'utilisent pas de LGRB intègrent leurs références manuellement dans un fichier formaté (58,46 %), ou intègrent directement les références dans le texte sans le concours du traitement de texte (29,17 %). Ils ne sont que 8,33 % de ceux qui n'utilisent pas de LGRB en sciences dures à utiliser les fonctionnalités dédiées d'un traitement de texte.

Intéressons-nous à la population de documentalistes et bibliothécaires. A la question « Utilisez-vous au moins un logiciel de gestion de références bibliographiques ? », la réponse a été quasi unanimement affirmative comme l'indiquent les résultats ci-après :

- utilise au moins un logiciel de gestion bibliographique : 96,43 %
- n'utilise aucun logiciel de gestion bibliographique : 3,57 %
- sans objet : 0 %

Nous allons nous intéresser aux logiciels qui sont utilisés au sein de la population. Parmi les logiciels signalés par les sondés sous la catégorie « autres », nous avons eu la surprise de voir des personnes utiliser le logiciel de « *concepts map* » XMind pour noter les références bibliographiques. Cette pratique peut surprendre, mais cela permet de visualiser les rapports de coécritures et de citations entre documents. Les autres logiciels proposés sont ReferenceManager (deux fois par des personnels de SCD) et le logiciel libre Pybliographer[13], une fois. La version Web d'EndNote a été également citée deux fois. Deux personnes en sciences dures, un enseignant chercheur et un post doctorant ont déclaré utiliser le portail ACM comme gestionnaire de bibliographie. Après vérification, le portail ACM propose une option soumise à abonnement « *My Binders* » pour stocker des hyperliens vers les notices des articles repérés sur le portail.

---

[13] Pour BibTeX : http://www.framasoft.net/article4108.html (accédé le 4 janvier 2012)

Deux personnes utilisent le portail de dépôt d'article en ligne du CNRS HAL pour gérer des notices bibliographiques.

Pour la répartition de l'utilisation des LGRB, de manière générale, le plug-in de Firefox Zotero est largement plébiscité. Parmi les progiciels, EndNote de Thomson Reuters, bien que payant, est également très largement utilisé (29,38 % de la population). Il arrive que les deux logiciels soient utilisés conjointement (12,84 % de l'ensemble du panel). Tous les pourcentages proposés pour la ventilation des usages de LGRB (Tableau 13 et Figure 4) s'entendent par rapport au nombre de personnes qui en utilisent dans l'échantillon et pas par rapport à la totalité de la population de référence.

En sciences humaines et sociales, Zotero est très largement usité (presque 63 % des utilisateurs de LGRB en SHS). Vient ensuite EndNote (presque 30 % des utilisateurs de LGRB en SHS) qui est parfois utilisé conjointement avec Zotero. Cela s'explique par le fait que les utilisateurs peuvent utiliser Zotero de deux manières. Ce plug-in peut être un logiciel unique dont l'usage va de la détection sur les sites compatibles de notices bibliographiques à l'export dans Word ou OpenOffice en passant par le stockage et la gestion de la bibliographie. Mais Zotero peut également être utilisé juste pour détecter les documents scientifiques. Les notices sont ensuite exportées dans EndNote qui se chargera et gestion et de l'intégration dans le traitement de texte.

L'écriture de documents en sciences dures nécessite souvent l'usage du format BibTeX, cela explique que JabRef arrive en tête des logiciels en terme d'utilisation avec presque 30 % des utilisateurs de LGRB en sciences dures. Mendeley (38,71 %) et Zotero (32,26 %) sont également largement utilisés. Zotero est d'ailleurs parfois utilisé en collaboration avec Mendeley et JabRef, pour les mêmes raisons et dans les mêmes conditions qu'en SHS. Mendeley offre la possibilité de communiquer avec Word grâce à un plug-in, alors que JabRef se spécialise dans la création et la gestion de fichiers BibTeX. JabRef par exemple permet de détecter les erreurs de typage dans un fichier BibTeX. Le chercheur s'épargnera ainsi une fastidieuse étape de débogage à la compilation de la bibliographie du document.

| LGRB utilisé | % global de la population | % en SHS | % en sc. Dures | % En SCD |
|---|---:|---:|---:|---:|
| Zotero | 54,13 % | 62,90 % | 32,26 % | 88,89 % |
| JabRef | 20,18 % | 9,68 % | 45,16 % | 14,81 % |
| RefWorks | 0,92 % | 1,61 % | 3,23 % | 0 % |
| BibDesk | 6,42 % | 3,22 % | 16,13 % | 0 % |
| Mendeley | 21,10 % | 11,29 % | 38,71 % | 25,93 % |
| EndNote | 29,36 % | 33,87 % | 16,13 % | 40,74 % |
| Bibus | 0,92 % | 0 % | 0 % | 3,70 % |
| RefBase | 0 % | 0 % | 0 % | 0 % |
| Autres | 11,01 % | 1,61 % | 12,90 % | 3,70 % |

Tableau 13. Ventilation de l'usage des LGRB

Zotero est très largement utilisé par les documentalistes et la seule personne en SCD de notre panel qui n'utilise pas de logiciel de gestion de bibliographie, se sert de notices Marc individuelles générées par le système de gestion de la bibliothèque. Dans l'ensemble, les personnels de SCD utilisent un ou plusieurs LGRB. Zotero est utilisé par presque 89 % des documentalistes et bibliothécaires dans l'enseignement supérieur. En complément et dans les mêmes conditions qu'en SHS ou Sciences dures, ils utilisent EndNote (40,74 %) ou Mendeley (presque 26 %).

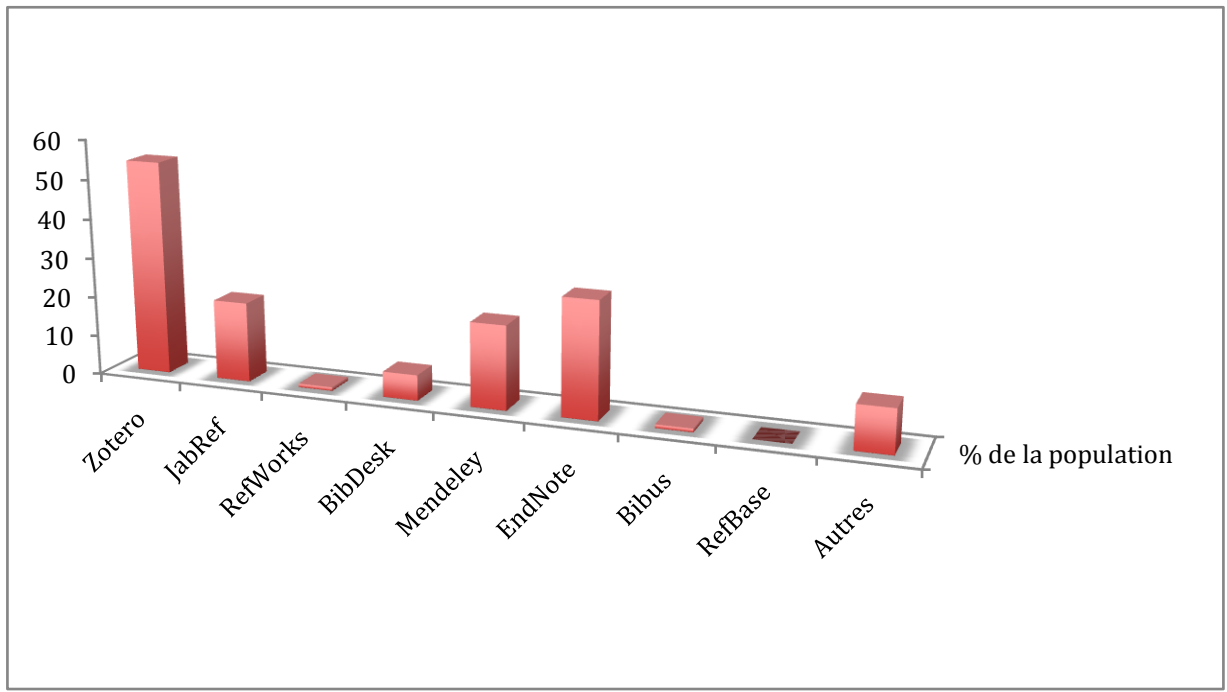

**Figure 6. Logiciels utilisés pour la gestion documentaire, parmi ceux qui en utilisent au moins un.**

### e) Critères pour le choix d'un LGRB

Les facteurs qui orientent le choix d'un logiciel de gestion de références bibliographiques sont multiples. Outre ceux d'ordre technique induits par le domaine de recherche, nous allons essayer d'en dégager plusieurs autres.

Dans un premier temps, intéressons-nous à la sensibilité aux logiciels libres ou Open Source qu'ils soient gratuits ou payants, opposés aux logiciels propriétaires qu'ils soient gratuits ou payants.

Sur l'ensemble de la population, 62,6 % préfèrent utiliser la première catégorie. 7,7 % des sondés se déclarent plus sensibles aux logiciels propriétaires. Enfin, 29,7 % n'ont pas de préférence par rapport à ce critère. Cette proportion est à peu de choses près équilibrée, quel que soit le profil.

Pour les doctorants en début de thèse, la donne est différente. Sur une population de 44 personnes, 4 (9,1 %) préfèrent les logiciels propriétaires, 17 (38,6 %) les logiciels libres et 23 (52,3 %) sont indifférents à ce critère. En fin de thèse nous assistons à une évolution : seuls 5,6 % continuent de préférer les logiciels propriétaires, 66,7 % se sont tournés vers le logiciel libre et 27,7 % n'ont pas d'opinion sur ce sujet. À l'étape suivante, pour les post doctorants et les enseignants chercheurs, le taux de préférence pour le logiciel propriétaire est 8,6 %. L'adhésion au logiciel libre monte à 72,4 % et l'absence de préférence descend à 20 %.

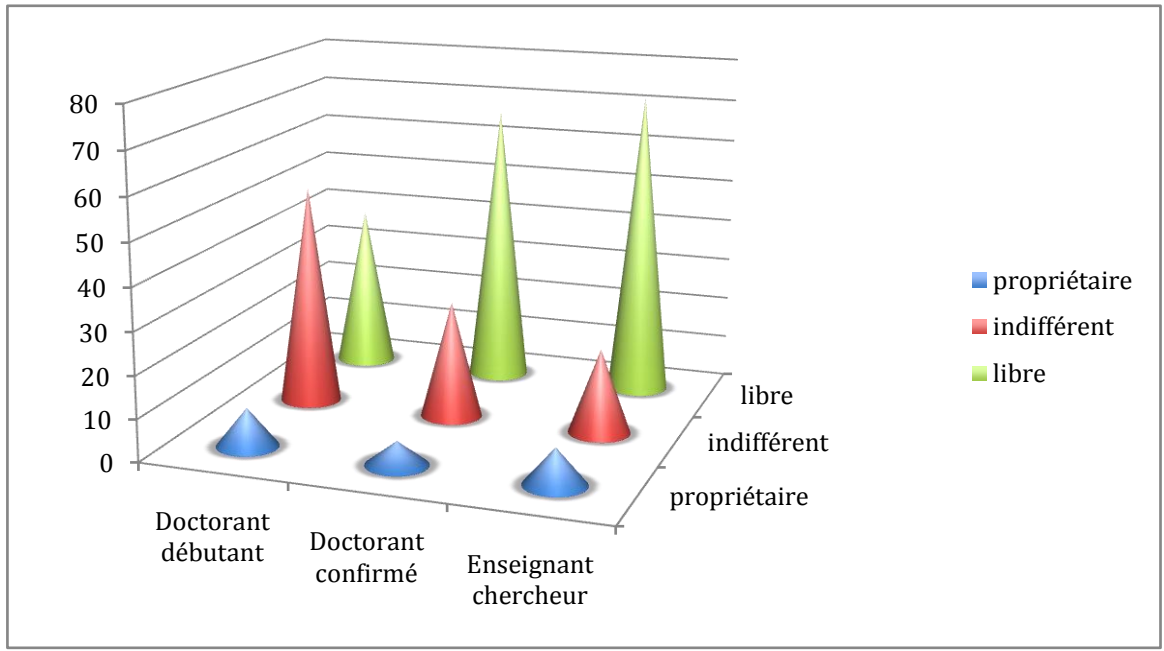

**Figure 7. Évolution de la sensibilité aux logiciels libres ou propriétaires (en % des groupes sondés) chez les chercheurs.**

La figure 7 illustre très clairement qu'avec l'expérience l'attachement au logiciel libre augmente – pas forcément au détriment des logiciels propriétaires – et l'indifférence au critère libre/propriétaire diminue. L'opinion se tranche donc, en faveur la plupart du temps des logiciels libres. Il est possible d'imputer le manque d'intérêt pour ce critère en début de doctorat à une phase d'observation pendant laquelle le jeune chercheur utilise ce dont il dispose. En fin de doctorat, il se sera imprégné des méthodes de son environnement et aura également appris à se servir des outils.

Après une première question générique sur les logiciels en général, la question suivante portait sur choix potentiel d'un logiciel de gestion bibliographique. Nous désirions savoir sur quels critères principaux s'oriente plutôt le choix des sondés. Le choix se portait sur les critères suivants : fonctionnalités du produit, facilité d'installation, gratuité, la disponibilité au sein de la structure (laboratoire, SCD), choix de vos collègues en la matière ou autre (avec invitation à préciser). Les sondés pouvaient choisir entre une et trois réponses ou ne pas se prononcer. De manière générale, 71 % des sondés s'intéressent avant tout aux fonctionnalités du produit, et ce avant la gratuité 57 %. Pour la moitié de la population, la facilité d'installation est un critère important. Notons à ce propos que parfois des logiciels libres nécessitent une compilation avec des dépendances logicielles. Cela les rend difficiles à installer pour des néophytes en informatique. Le choix des collègues et la disponibilité dans la structure sont des critères importants pour 20 % des sondés. Comme propositions alternatives, nous avons eu : « la facilité d'utilisation » et « l'habitude d'usage ».

Notons à ce propos que parmi ceux qui disposent d'un LGRB, 78,1 % l'ont choisi et installé eux-mêmes, 14,5 % utilisent celui proposé par l'organisme de rattachement déployé par défaut (EndNote dans 62,5 % des cas). L'appartenance à un groupe scientifique ou une catégorisation personnel/doctorant n'est pas un facteur discriminant dans ce cas. Cela mène à penser qu'il s'agit de licences établissement négociées à grande échelle et donc installées aussi bien en laboratoire qu'en centre de documentation.

Pour les 7,5 % restant, ils ont dû installer eux-mêmes le logiciel choisi par l'organisme de rattachement (EndNote à 50 %, installation mixte de Zotero et d'EndNote à 50 %).

Ceux qui ont choisi et installé eux-mêmes leur LGRB ont le plus souvent choisi Zotero (38 %), souvent associé avec un autre logiciel comme Mendeley, EndNote ou JabRef. Cela tend à confirmer que Zotero est surtout utilisé pour le glanage d'informations scientifiques, mais que la gestion de la bibliographie est plutôt confiée à un logiciel tiers dédié à cet usage. En effet, seuls 20 % de la tranche que nous étudions utilisent Zotero sans logiciel complémentaire. Mendeley, EndNote et JabRef sont installés et utilisés par 15 % d'utilisateurs chacun.

**f) Recherche d'informations**

Pour la recherche d'informations, les sondés devaient spécifier entre une et trois sources spécifiques parmi les suivantes :

- Les OPAC[14] des SCD
- Sur le SUDOC
- Les moteurs scientifiques (Google Scholar, Scirus…)
- Les sites d'éditeurs scientifiques (Elsevier, Springer, IEEE, ACM…)
- Les archives ouvertes (HAL, ArXiV, Archivesic…)
- Les moteurs de recherche traditionnels (Google, Bing ou Yahoo…).

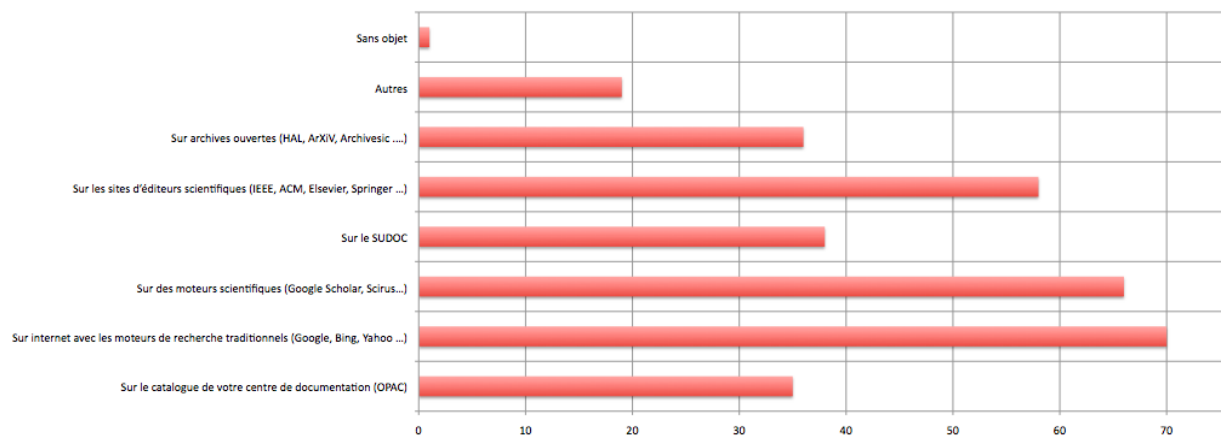

**Figure 8. Sources d'informations scientifiques pour la recherche dans l'enseignement supérieur.**

Nous observons en premier lieu que pour la recherche d'informations scientifiques et techniques, dans l'enseignement supérieur la source première d'information reste les moteurs de recherche commerciaux traditionnels. Cette observation n'est pas nouvelle, elle avait déjà été observée dans des études américaines. Si l'on en croit Markey, les catalogues des bibliothèques sont tombés en disgrâce [20], c'est ce que semble confirmer notre étude. Notre étude montre que, sur un Panel de 195 doctorants, enseignants-chercheurs et personnel de SCD français, la recherche d'informations scientifiques et techniques ne se fait sur les OPAC que pour 35 % des publics sondés. Une étude de menée en 2006 pour le compte de l'Online Computer Library Center a démontré que rares sont les étudiants, enseignants et chercheurs commençant leurs recherches par le catalogue de leur bibliothèque [21]. Ils sont 89 % à lui préférer un moteur de recherche commercial, au premier rang desquels Google. Selon l'étude de 2005 [22],72% des universitaires utilisent le moteur Google. Lorcan Dempsey, directeur de la recherche à OCLC[15], a qualifié notre époque d'ère « *Amazoogle* » faisant allusion au moteur de recherche Google et à la librairie numérique Amazon.

---

[14] OPAC – Open Access Catalog, catalogue de bibliothèque accessible depuis internet.

Nous allons analyser nos chiffres sur cette question par population. Les jeunes doctorants utilisent leur OPAC pour un tiers d'entre eux. En fin de thèse, ce chiffre chute à 28 %. La surprise vient de la distinction des usages selon la science étudiée.

1. L'utilisation des sources en sciences dures

En sciences dures, les doctorants n'utilisent pas (du tout) le catalogue en ligne de leur SCD. Nous avons assimilé les post doctorants aux enseignants-chercheurs, du fait de leur faible nombre dans l'étude, pour les chiffres suivants. Seuls 22 % de la population utilise l'OPAC de son établissement pour accéder à l'IST. Encore une fois, la proportion est plus faible en sciences dures avec 7,5 %, là ou en SHS notre étude en dénombre 78,6 %.

En sciences dures, la recherche d'IST se fait davantage au travers de moteurs de recherche scientifique comme Google Scholar ou Scirus (78 % des sondés). La population de sciences dures fait usage des éditeurs scientifiques dans les mêmes proportions (76 %). Les moteurs traditionnels comme Google, Yahoo ou Bing sont également largement utilisés (71 %).

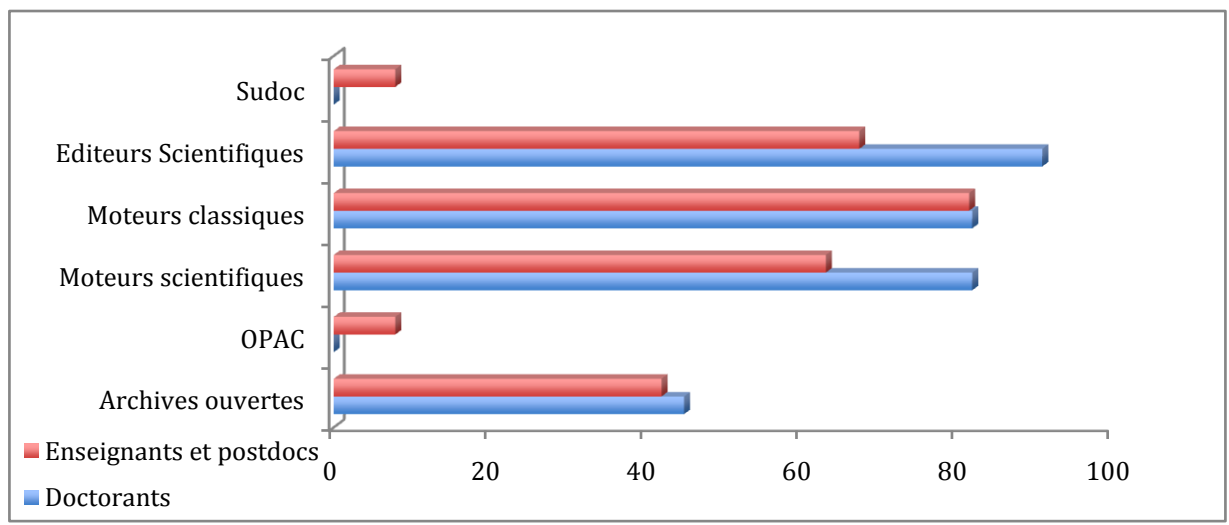

Figure 9. Sources documentaires en sciences dures consultées par profil d'expérience.

2. L'utilisation des sources en sciences humaines et sociales

En sciences humaines et sociales, nous avons pu distinguer trois populations différentes du fait d'un plus grand échantillon. Les pratiques en recherche d'informations scientifiques et techniques (IST) évoluent avec l'expérience des usagers. En début de thèse, les trois principales sources électroniques d'IST sont les moteurs de recherche scientifique et classiques, mais aussi l'OPAC de l'établissement. Pour ce qui est de l'OPAC, les doctorants sont 44 % à l'utiliser en début de thèse et 33 % en fin de thèse. En fin de thèse, les moteurs traditionnels sont en tête d'utilisation, puis viennent le SUDOC et les moteurs scientifiques. Ces derniers arrivent à égalité avec les archives ouvertes.

---

[15] *L'Online Computer Library Center* (OCLC) est une organisation de services, d'études et de développement à l'attention des bibliothèques du monde entier.

Enfin, les enseignants chercheurs utilisent aussi bien les moteurs scientifiques et classiques que le SUDOC en premier lieu. Ils utilisent moins les OPAC (44 %), ce qui reste néanmoins très au-dessus de la moyenne générale d'utilisation des OPAC.

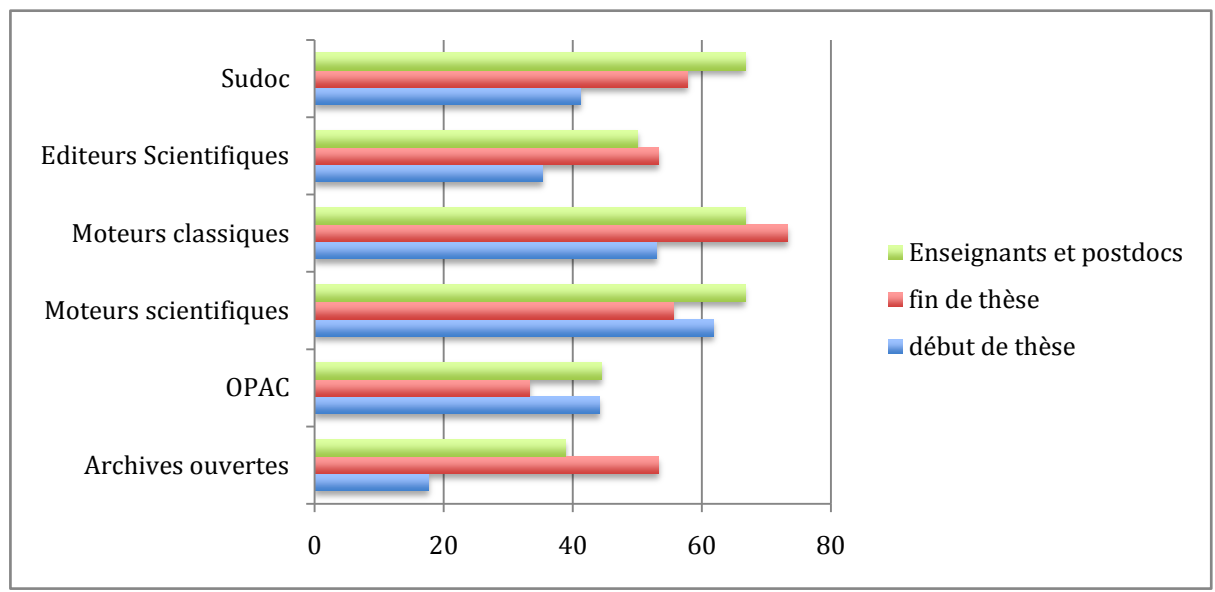

Figure 10. Sources documentaires consultées en sciences humaines par profil d'expérience.

Il faut retenir de ces quelques données que les chercheurs et étudiants chercheurs en sciences humaines usent volontiers de leurs OPAC. En sciences dures, ce n'est pas le cas.

3. Utilisation de sources pour les documentalistes et bibliothécaires

En SCD, la situation est bien différente puisque 78,6 % de l'échantillon se sert de son OPAC. Nous pouvons émettre l'hypothèse que les personnels de SCD maîtrisent parfaitement le catalogue en ligne de l'établissement pour lequel ils travaillent. Ce n'est cependant pas leur seule source d'IST.

Les documentalistes et bibliothécaires ont également une bonne pratique des éditeurs scientifiques, 71,5 % d'entre eux les utilisent pour la recherche d'IST. D'ailleurs, les centres documentaires sont souvent abonnés à plusieurs éditeurs scientifiques, ce qui représente un axe de dépense non négligeable. Les moteurs de recherche classique (60 % d'utilisation en SCD) ou scientifique (50% d'utilisation) ne sont pas les sources d'informations privilégiées par les personnels de documentation interrogés. Cela s'explique assez facilement par le fait que l'information est difficilement vérifiable et mal formatée sur un moteur de recherche standard. Pour les moteurs de recherche scientifique, au premier rang desquels émerge Google Scholar, la fiabilité des données et métadonnées est trop contestable pour être appréciée par un documentaliste[16]. Nous avons noté au cours de nos essais des fonctionnalités de Google Scholar que le formatage des données bibliographiques est approximatif. Les données elles-mêmes sont partielles, ce qui les rend inexploitables en l'état. Il faut notamment régulièrement réajuster le type du document. Le type « *Inproceedings* » (actes de conférence) par exemple est régulièrement remplacé par le générique « *article* ».

---

[16] Lardy disait de Google Scholar : « Google Scholar est un bon point de départ mais qu'il n'a pas encore la maturité des outils de recherche documentaires commerciaux [23].»

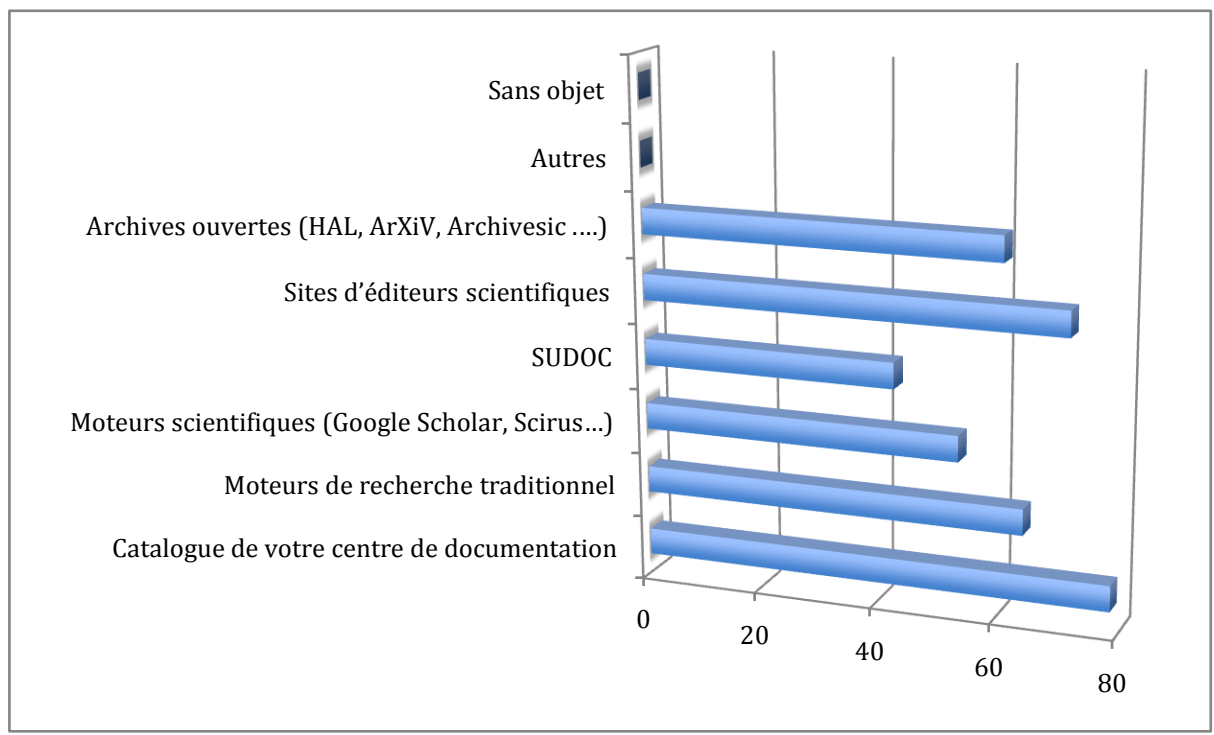

Figure 11. Sources de données en centre de documentation

Avec 60 % d'utilisation en SCD, les archives ouvertes sont moyennement utilisées par les documentalistes et bibliothécaires. Les données sont souvent redondantes entre archives ouvertes et éditeurs scientifiques. Parfois, sur les archives ouvertes ne sont saisies que des notices bibliographiques. Ces notices ne sont parfois pas saisies par les auteurs [24], d'où des inexactitudes dans les métadonnées. Comme les notices sont automatiquement répliquées d'une archive à l'autre, les potentielles erreurs sont multipliées. Cela explique peut-être, la plus faible utilisation des archives ouvertes chez la catégorie des documentalistes et bibliothécaires, très soucieux de la cohésion des données récoltées.

Le SUDOC est assez peu utilisé (40 % des professionnels de la documentation sondés). Il est peut être consulté en dernier recours, dans l'optique de localiser un ouvrage rare et demander un prêt inter établissement du document primaire.

Pour la plupart d'entre eux, les personnels de centres de documentation utilisent le plus large panel de sources documentaires des membres de la communauté de l'enseignement supérieur.

**Analyse des résultats**

Grâce à l'interprétation de cette collection de données, nous allons tenter d'établir des profils de pratiques techniques d'utilisateurs de systèmes d'informations documentaires. Ces profils permettront ultérieurement de définir un cahier des charges pour la modélisation de SRI par population.

**a) Profil Sciences humaines et sociales**

En SHS, le profil technique type de l'utilisateur de SRI est composé d'un environnement équipé d'un système d'exploitation Windows. Les doctorants sont majoritairement équipés de Windows, cette tendance évolue après le doctorat avec un usage équilibré entre Mac OS et Windows. L'outil rédactionnel est très majoritairement le traitement de texte avec un petit pourcentage qui s'oriente vers le compilateur avec l'expérience.

Pour ce qui est des usages typiquement liés à la bibliographie, une moitié intègre les citations et rédige sa bibliographie manuellement, ce qui est en accord avec le fait que moins d'un sondé en SHS sur deux connaisse au moins un format de fichier bibliographique. Néanmoins, bien que peu utilisé, le format BibTeX est connu par un quart de l'échantillon SHS. Les format XML OpenDocument et RIS sont utilisés chacun par presque 10 % de l'échantillon SHS. Pour la gestion de la bibliographie, la moitié des sondés utilise une assistance logicielle dédiée à cet usage (LGRB). Parmi ceux qui n'en utilisent pas, un sur cinq se sert des fonctionnalités intégrées au traitement de texte. Mais, le plus souvent, les éléments bibliographiques sont intégrés et gérés sans assistance logicielle dans les documents.

Pour ceux qui utilisent un LGRB en sciences humaines et sociales, la pratique la plus courante est d'utiliser Zotero pour la détection de notices et de l'associer à Mendeley ou EndNote.

En SHS, des sources multiples sont largement utilisées, les moteurs classiques et scientifiques bien sûr, mais aussi les éditeurs scientifiques. Les OPAC sont, en revanche, moins utilisés, mais beaucoup plus qu'en sciences dures. Les archives ouvertes sont assez peu utilisées, sauf chez les doctorants en fin de thèse.

**b) Profil sciences dures**
En sciences dures, pour les chercheurs, le choix des systèmes d'exploitation est plus hétérogène qu'en SHS. Les systèmes basés sur Unix comme Linux ou Mac OS sont privilégiés dès la fin de la thèse, le temps que les doctorants s'en approprient l'usage. Cet usage trouve probablement son explication dans la quantité d'outils scientifiques disponibles sous ces systèmes d'exploitation. Outils au rang desquels les compilateurs de texte de type TeX. Ces outils sont largement utilisés pour la production de documents scientifiques dès la fin de la thèse. Cela induit le choix de BibTeX comme format de bibliographie dans la plupart des cas.

Pour gérer leur bibliographie, les chercheurs en sciences dures utilisent pour plus de la moitié d'entre eux un logiciel dédié, souvent JabRef ou Mendeley en collaboration avec Zotero. Encore une fois, un usage conjoint de Zotero avec un autre outil permet de repérer et d'enregistrer les notices bibliographiques depuis internet puis de gérer la bibliographie avec un logiciel dédié.

L'usage de moteurs de recherche classique ou scientifique est très répandu pour la recherche d'IST en sciences dures, de même que les portails en ligne des éditeurs scientifiques.

**c) Le profil documentaliste (ou bibliothécaire) en SCD.**
Les personnels dans les SCD de sciences dures sont équipés de systèmes Windows et de Mac OS pour ceux spécialisés en sciences humaines. Dans leur ensemble, les documentalistes et bibliothécaires de l'enseignement supérieur utilisent exclusivement un traitement de texte. Ils connaissent et utilisent majoritairement les formats BibTeX et RIS. Ils génèrent les fichiers à ces formats depuis Zotero, puis les gèrent avec EndNote ou Mendeley. Pour la recherche électronique d'IST, cette population use d'un large panel de sources. Parmi ces sources, l'OPAC de leur établissement occupe une bonne place. Les éditeurs scientifiques font également partie de leurs sources favorites.

## Conclusion et perspectives

Selon le sociologue français Vourc'h, le pourcentage d'étudiants prétendant se rendre à la bibliothèque universitaire au moins une fois par semaine a baissé de 54 % en 1997 à 49,9 % en 2006 [25]. Pour ce qui est des outils en ligne, l'usage des OPAC est quasi nul pour les doctorants en sciences dures et faible pour les enseignants chercheurs, plus habitués aux moteurs traditionnels et scientifiques, mais aussi aux éditeurs en lignes. En sciences humaines, la consultation de l'OPAC est plus élevée, mais pas au même niveau que les moteurs classiques et scientifiques. Pourtant, le catalogue en ligne des centres documentaires semble parfaitement convenir aux documentalistes et bibliothécaires qui maîtrisent parfaitement leurs arcanes, mais aussi les outils, méthodes et formats bibliographiques.

Comme le montrait le dossier « Web sémantique, web de données » de la revue Documentaliste - Sciences de l'information de décembre 2011, pour mieux couvrir les besoins technologiques en matière informationnelle, il est possible d'envisager de faire des points d'accès multiples à même information scientifique et technique comme pour DBLP[17] ou ISIDORE[18] [29]. De plus, les autres facteurs inhérents au web de données doivent être respectés comme la compatibilité inter logicielle, l'exposition des métadonnées [30]. Selon l'axe de lecture scientifique et les choix techniques en matière de pratique bibliographique, les utilisateurs pourraient alors avoir un accès adapté à leurs attentes.

Il est évident qu'une bonne connaissance des outils normes et formats utilisés par les usagers de l'information est un pré requis pour modéliser un SRI compatible avec la population cible. On n'envisagera ainsi pas un outil de recherche d'IST en sciences humaines de la même manière qu'un SRI en sciences dures. Cependant, la constante dans tous les cas est que l'information se doit d'être visible et exportable, mais aussi compatible avec les outils de glanage contextuel d'information comme Zotero.

## Remerciements

---

[17] DBLP offre une interface classique « *human readable* » et une interface d'interrogation SparQL [26] qui connaît un vif succès parmi le monde la recherche [27] [28]
[18] Isidore propose trois interfaces : http://rechercheisidore.fr/ pour l'interface de recherche intuitive, http://rechercheisidore.fr/api pour les interconnections avec les outils logiciels et  http://rechercheisidore.fr/sparql pour les adeptes du langage de requête SparQL [26].
[19] Institut National des Techniques de la Documentation

## Bibliographie